\documentclass[12pt]{JHEP3} 

\usepackage{amsmath,amssymb,axodraw}

\newcommand\fverb{\setbox\pippobox=\hbox\bgroup\verb}
\newcommand\fverbdo{\egroup\medskip\noindent%
                        \fbox{\unhbox\pippobox}\ }
\newcommand\fverbit{\egroup\item[\fbox{\unhbox\pippobox}]}
\newbox\pippobox

\title{Dynamical $\mu$ and MSSM}

\author{Jihn E. Kim
         \\
          School of Physics and Center for Theoretical Physics,
        Seoul National University, Seoul 151-747, Korea \\
        E-mail: \email{
        jekim@phyp.snu.ac.kr} }

\preprint{SNUTP 05-002}      

\abstract{We present the idea that the vacuum can choose one pair
of Higgs doublets by making the $\mu$ parameter a dynamical field
called {\it massion}. The {\it massion} potential leading to the
dynamical solution is suggested to arise from the small instanton
interaction when the gauge couplings become strong near the cutoff
scale $M_s$ or $M_P$. One can construct supergravity models along
this line. We also present an explicit example with a
trinification model from superstring.}


\begin{document}


\def\lsl{ l \hspace{-0.45 em}/}
\def\ksl{ k \hspace{-0.45 em}/}
\def\qsl{ q \hspace{-0.45 em}/}
\def\psl{ p \hspace{-0.45 em}/}
\def\ppsl{ p' \hspace{-0.70 em}/}
\def\dsl{ \partial \hspace{-0.45 em}/}
\def\Dsl{ D \hspace{-0.65 em}/}
\def\dsl{ \partial \hspace{-0.55 em}/}
\def\matrix{ \left(\begin{array} \end{array} \right) }
\def\H{{\cal H}}
\def\MGUT{M$_{\rm GUT}$}

\baselineskip 0.7cm

\section{Introduction}

It is widely believed that  the minimal supersymmetric standard
model(MSSM) is the most probable immediate extension beyond the
standard model(SM). It has three families of quarks and leptons,
the SM gauge bosons, and their superpartners, and {\it one pair}
of Higgs doublet superfields $\H_1$ and $\H_2$. The MSSM problem
of obtaining just one pair is more constrained than the $\mu$
problem \cite{mu} or the doublet-triplet splitting problem
\cite{dimgeo}. For example, in the $Z_3$ orbifold compactification
one can easily realize the doublet-triplet splitting \cite{iknq},
but the minimum number of doublets is three pairs. In this paper,
we look for a possibility of dynamical solution of the MSSM
problem, by promoting $\mu$ as a dynamical field.

The most well-known dynamical solution of a coupling constant is
the axion solution of $\theta$ parameter \cite{PQVW}. In the
$\theta$ vacuum, the Euclidian space partition function determines
the vacuum energy $E(\theta)$ as,
\begin{align}
e^{-VE(\theta)}= \left|\int dA^a_\mu \ {\rm det}(\Dsl+M)\exp
\left[ -\frac{1}{4g^2}\int d^4x {\rm Tr} F_{\mu\nu}  F_{\mu\nu}
\right] \exp\left(\frac{i\theta}{16\pi^2}\int d^4x {\rm Tr}F\tilde
F\right)\right|.\label{ZVaWi}
\end{align}
From Eq. (\ref{ZVaWi}), one can show by using the Schwarz
inequality that $E(0)\le E(\theta)$. This  is the basis of the
Peccei-Quinn mechanism making the vacuum choose $\theta\equiv
a/F_a=0$. For this mechanism to work, at tree level there is no
potential of the axion field, i.e. it is a flat direction at tree
level. The axion potential comes only from the one loop correction
of the anomaly term.

Here, we ask a similar question on the $\mu$ term whether it can
be understood dynamically. Actually, in supersymmetric models the
determinental factor contains $(\Dsl+M_f)^2/(D^2+M_b^2)$,
revealing  the information on the potential of the flat
directional real scalar fields if it appears in the mass matrix.
Thus, the mass matrix $M$ in spontaneously broken supersymmetric
models can be used for this purpose. We require that $\mu\equiv
s$  does not have a potential at tree level so that $\mu$ can
become a dynamical field {\it \`a la} the axion solution. Then,
the effective potential with one-loop correction can be taken as
\begin{align}
V=V_{\rm 0} +\frac{1}{64\pi^2}{\rm Tr}\ (-1)^FM^4\ln
\frac{M^2}{\lambda^2}\label{Oneloop}
\end{align}
where $V_0$ is the tree value and $\lambda$ is the renormalization
scale.\footnote{We do not use the customary renormalization scale
$\mu$ to avoid the confusion with the $\mu$ term.} If
supersymmetry is not broken, in the vacuum $V_0=0$ (in the global
limit) and the one-loop correction vanishes. This has the needed
property of the flat direction for $s$. This flat direction is
{\it massion}, named for its role of determining the mass
parameter of the Higgs doublets. This flat direction is lifted
once supersymmetry is softly broken. Expressing the generic
magnitude for the soft supersymmetry breaking as $\delta^2$, the
flat-direction lifting term is of order $\delta^2M_P^2$.

However, the form (\ref{Oneloop}) is not the one we expect toward
a pair of light Higgs doublets since the minima generally do not
choose a massless doublet. If the potential is of the form
Det.$M_f$ as suggested in \cite{cchk}, then a massless Higgsino
doublet(s) and hence a massless pair(s) of Higgs bosons will
follow. A possibility for this kind of determinental interaction
is present if small scale instantons are important. Naively, one
would expect that the small scale instantons would not affect the
low energy physics significantly. But due to the possibility of
packing a large number of instantons within a given volume if the
instanton size is small (if the gauge coupling becomes strong at
high energy), i.e. from the instanton size integration $\int
d\rho/\rho^5$, the small scale instanton contribution to a small
physical parameters can be significant. Indeed, the contribution
of small scale QCD instantons was considered to the axion
potential if QCD becomes strong at very high energy scale
\cite{choihd}. Of course, small scale instantons of other
nonabelian groups can be important to the potential of almost flat
directions. In this paper, we study such a possibility toward the
potential of Higgs doublet fields.

We find that the useful small scale instantons toward the MSSM is
the $q=4$ instanton of the diagonal subgroup of
SU(2)$_R\times$SU(2)$_L$ where SU(2)$_L$ is the electroweak SU(2).
This kind of embedding is possible in the trinification type and
Pati-Salam type models.

In Sec. \ref{Sec:OneLoop}, we show that the one loop potential
(\ref{Oneloop}) is of order $M_W^2M_P^2$ and does not give a
massless doublet.  In Sec. \ref{Sec:dynfield}, we introduce a
relatively strong force at a high energy scale. In Sec.
\ref{Sec:tangled}, we present the tangled instanton  which does
not emit ordinary quarks and leptons but  emits Higgsinos. In Sec.
\ref{Sec:tri}, we present this idea in a trinification model. Sec.
\ref{Sec:conclusion} is a conclusion.

\section{One loop potential}\label{Sec:OneLoop}

\begin{figure}[h]
\begin{center}
\begin{picture}(400,200)(0,0)
\LongArrow(60,20)(320,20)\Text(330,23)[l]{$m^2$}
\LongArrow(80,0)(80,190)\Text(70,180)[c]{$\tilde V$}
\Curve{(80,24)(110,30)(300,180)}
\end{picture}
\caption{One pair of Higgs doublets.}\label{Onepair}
\end{center}
\end{figure}
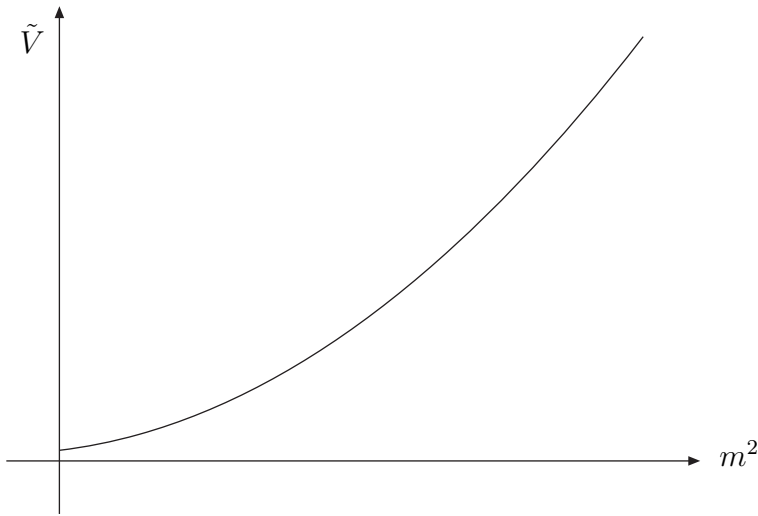

Before introducing the massion $s$, let us consider the mass
matrix $M$ of (\ref{Oneloop}).  For one pair of chiral multiplets,
$S$ and $\bar S$, with a common mass splitting of $\delta^2$, the
effective potential is
\begin{align}
V_1=V_{\rm 0} +\frac{2}{64\pi^2}\left[
(m^2+\delta^2)^2\ln\frac{(m^2+\delta^2)}{\lambda^2} -
m^4\ln\frac{m^2}{\lambda^2}\right]\equiv V_{\rm 0}
+\frac{2}{64\pi^2} \tilde V_1\label{exponeloop}
\end{align}
where
\begin{equation}
\tilde V_1=\left\{\begin{array}{l}
m^4\ln\left(1+\frac{\delta^2}{m^2}\right)+(2m^2\delta^2+\delta^4)
\ln\frac{m^2+\delta^2}{\lambda^2},\ \ {\rm for\ }m^2\ne 0,\ \
m^2>\lambda^2\\
\delta^4\ln\frac{\delta^2}{\lambda^2},\ \ {\rm for\ }m^2=0,\ \
\delta^2>\lambda^2
\end{array}
\right.\label{ScaleLam}
\end{equation}
Thus, the magnitude of $V_1$ is of order $\delta^2$. If we make
$m^2$ a dynamical variable, we can compare $\tilde V(m^2=0)$ with
other values of $\tilde V$.  We take the renormalization scale
$\lambda^2$ less than $m^2+\delta^2$  so that the bosonic
contribution is positive. The shape of $\tilde V$ has the $m^2$
dependence as shown in Fig. \ref{Onepair}. In this case, $m^2=0$
is the minimum of the one-loop potential. However, this property
does not persist if there exist more than one pair of Higgs
doublets.

We are interested in the following range of parameters,
\begin{align}
\delta^2\sim {\rm TeV}^2,\ m^2\sim M^2_P,\ m_{3/2}\sim {\rm TeV}
\end{align}
where $\delta^2$ is the mass splitting in spontaneously broken
supergravity. Then, the $A$-term in $V_0$  has the contribution
$$
AmH_1H_2\to m_{3/2}v^2 m
$$
where $v$ is the electroweak scale VEV. Thus, the $A$-term is
negligible compared to $\delta^2m^2$.

\begin{figure}[t]
\begin{center}
\begin{picture}(400,110)(0,0)
\LongArrow(60,20)(320,20)\Text(330,22)[l]{$b$}
\LongArrow(200,0)(200,110)\Text(190,100)[c]{$\tilde V$}
\Curve{(80,90)(190,24)(300,100)}
\Line(250,18)(250,25)\Text(250,10)[c]{$a$} \Text(100,10)[c]{$-2a$}
\Line(100,18)(100,25)
\end{picture}
\caption{More than one pair of Higgs doublets.}\label{Threepairs}
\end{center}
\end{figure}
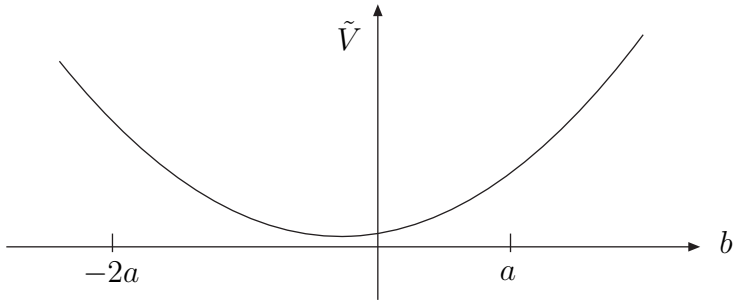

Now let us proceed to discuss a Higgsino mass matrix with an $S_3$
symmetry. One such matrix is
\begin{equation}
M_{\tilde H}=\left(
\begin{array}{ccc}
b\ & a\ & a\\
a\ & b\ & a\\
a\ & a\ & b
\end{array}
\right)\label{MMHiggsino}
\end{equation}
which has eigenvalues of
\begin{equation}
m_{\tilde H}=b+2 a,\ \ b-a,\ \ b-a.
\end{equation}
For this case, the shape of one-loop potential  looks like Fig.
\ref{Threepairs}.
With the mass splitting parameter $\delta^2$, the above potential
is shifted up by O($\delta^2 M^2$) where $M^2$ is the mass
parameter in the tree level potential, presumably of order the
Planck scale. In supergravity, the vacuum value can be fine-tuned
at $V=0$ \cite{WZgr}. Let us take the sign of $a>0.$ In Fig.
\ref{Threepairs}, $a$ and $\delta$ are considered as fixed numbers
and $b$ is considered as a variable parameter, and we assumed
$a^2\gg\delta^2$ where $a^2=O(M^2)$. Certainly, the minimum
position in Fig. \ref{Threepairs} is not near $b=a$ or $b=-2a$.
Thus, even if massion is introduced, the one loop potential of Eq.
(\ref{exponeloop}) is not guaranteeing a massless pair of Higgs
doublets. One needs another interaction for massion to choose a
massless pair of Higgs doublets.

\section{The $\mu$ term as a field}\label{Sec:dynfield}

\subsection{The $\mu$ problem}
The $\mu$ problem, ``Why is $\mu$ so small compared to the GUT
scale?", is a part of the MSSM problem for obtaining the MSSM
spectrum. The $\mu$ term is the mass term for a vectorlike pair of
fermions in the superpotential, i.e.
$$
W_{\mu_X}=\mu_X \bar X X
$$
where $X$ and $\bar X$ are left-handed chiral superfields and each
carries the charge conjugated  gauge quantum numbers of the other.
Thus, considering the gauge symmetry only, a non-vanishing $\mu_X$
is allowed and its magnitude is typically of order where the
theory is written. In the MSSM spectrum  at the electroweak scale,
there exists a vectorlike pair of Higgsinos, $\tilde H_1$ and
$\tilde H_2$. Thus, we expect a large $\mu_H$, presumably at the
GUT scale.  But, there is a need to obtain one pair of light
Higgsinos for the MSSM spectrum. This is the MSSM problem. If the
MSSM results from a GUT, then a similar term $\mu_T$ would appear
for a pair of vectorlike color triplet Higgsinos $\tilde T$ and
$\tilde{\bar T}$. But the MSSM does not need the color triplet
Higgsinos, which is called the doublet-triplet splitting problem.
Thus, {\it the MSSM problem is the problem of obtaining  a small
$\mu_H$ only for one pair of Higgsinos but keeping the other
$\mu_H$'s and all the $\mu_T$'s to remain large.}

Earlier suggestions for the solution of $\mu$ problem are firstly
using some symmetries to forbid the $\mu_H$ term of $\H_1$ and
$\H_2$ at the scale in consideration (but allow the $\mu_T$ term
for color triplet Higgsinos at high energy scale) \cite{musym}. In
particular, it should be forbidden in the superpotential. But, we
need the $\mu_H$ term of order the electroweak scale to obtain a
phenomenologically viable electroweak symmetry breaking. This is
achieved by breaking the Peccei-Quinn(PQ) symmetry at the
intermediate scale as suggested or at the scale leading to the
gravitino mass \cite{mu}. Since the PQ symmetry breaking scale and
the intermediate scale for supergravity are of the similar order,
both of them give reasonable electroweak symmetry breaking.
However, these suggestions do not give a rationale why only one
pair of Higgs doublets remains light.

\subsection{$\mu$ as a dynamical field}
On the other hand, recently an ansatz was suggested so that that
the MSSM problem can be understood dynamically \cite{cchk}. In
this spirit, we promote the {\it $\mu$ term as a dynamical field}
\cite{cchk}. The dynamical $\mu$ is called {\it massion}. For a
vectorlike representation, the group singlet field(s) can
contribute to the $\mu_X$ term, viz. $(\mu_X+s)\tilde X\tilde{\bar
X}$, and hence we will use  the mass parameter and the massion
field $s$ interchangeably.

As discussed in Sec. \ref{Sec:OneLoop}, if there is no other
contribution to the massion potential, then it is impossible to
obtain a massless pair of Higgs doublets from the extremum of the
one-loop potential for the massion field. Therefore, we need a
relatively strong force for this purpose.  In fact, we observe
that there exists such a possibility due to a large number of
matter fields allowable above the GUT scale \MGUT. A large number
of matter fields destroys the asymptotic freedom above \MGUT, and
gauge couplings become stronger going above \MGUT.
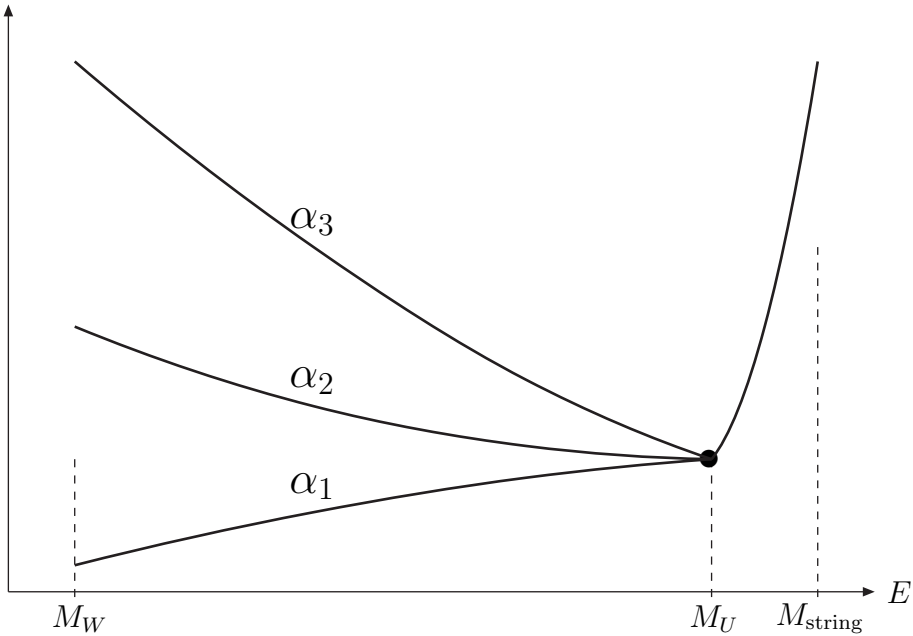
\begin{figure}[t]
\begin{center}
\begin{picture}(400,240)(0,-10)

\LongArrow(5,0)(330,0)\Text(340,0)[c]{ $E$} \LongArrow(5,0)(5,220)
\DashLine(270,50)(270,-2){3} \DashLine(310,130)(310,-2){3}

\Text(310,-10)[c]{ $M_{\rm string}$} \Text(270,-10)[c]{ $M_U$}
\DashLine(30,50)(30,-2){3} \Text(30,-10)[c]{ $M_W$}
\Text(270,50)[c]{\Large $\bullet$} \Text(120,140)[c]{\Large
$\alpha_3$} \Text(120,80)[c]{\Large $\alpha_2$}
\Text(120,40)[c]{\Large $\alpha_1$} \SetWidth{1.0}
\Curve{(30,200)(120,130)(200,80)(270,50)}
\Curve{(30,100)(120,70)(200,55)(270,50)}
\Curve{(30,10)(120,30)(200,43)(270,50)}

\Curve{(270,50)(290,100)(310,200)}

\end{picture}
\caption{Gauge couplings above the GUT scale.}\label{couplingrun}
\end{center}
\end{figure}
Neglecting Yukawa couplings, its behavior is shown in Fig.
\ref{couplingrun}.  We anticipate a situation that the nonabelian
scale $\Lambda$ where the interaction becomes strong is roughly
the fundamental scale(or string scale) so that a perturbative
discussion below the fundamental scale is possible.

\subsection{Small-instanton generated
potential}\label{Subsec:Actsmall}

If the nonabelian scale $\Lambda$ is around or above the string
scale, the field theory calculation of the nonperturbative effect
below the string scale is possible. In particular, the
small-instanton solution, which is small in our TeV scale jargon
but large at its nonabelian scale $\Lambda\ge M_s$, has the
amplitude proportional to $e^{-8\pi^2/g^2(\lambda)}$ with
$\lambda<M_s$. We will look for the situation where this
nonperturbative effect is effective in determining a small
parameter of the MSSM, i.e. the Higgsino mass parameter.

Before considering the asymptotically strong case, $N_f>3N_c$, let
us recapitulate the case $N_f<N_c$.

\subsubsection{Case $N_f<N_c$}
The nonabelian scale $\Lambda$ is best understood in
asymptotically free gauge models. So, for a moment consider
asymptotically free nonabelian gauge theories, before we propose
the coupling behavior of Fig. \ref{couplingrun} above \MGUT.
Roughly, it is the scale where the coupling becomes
strong.\footnote{For QCD, $\Lambda_{\rm QCD}$ is a few hundred
MeV.} Below the scale $\Lambda$, a QCD-like theory will show the
confinement and chiral symmetry breaking. In {\it asymptotically
free supersymmetric} QCD of SU($N_c$), let us consider $N_f$
pairs(or flavors) of left-handed superfields $Q$ and $\bar Q$ with
$N_f<N_c$. The classical Lagrangian has the following global
symmetries
\begin{align}
SU(N_f)_L\times SU(N_f)_R\times U(1)_A\times U(1)_B\times U(1)_R
\label{glosym}
\end{align}
with the following global quantum numbers of squarks and gauginos
\cite{IntSeib},
\begin{equation}
\begin{array}{cccccc}
Q\ :& (N_f, & 1, & 1, &1, &(N_f-N_c)/N_f)\\
\bar Q\  :& (1,& N_f, & 1,& -1,&(N_f-N_c)/N_f)\\
{\rm gauginos}\  :& (0,& 0, & 0,& 0,&+1).
\end{array}
\end{equation}
The U(1)$_R$ quantum numbers are chosen so that it is anomaly
free. The anomalous U(1) is just U(1)$_A$. At a long distance
limit, much larger than $\Lambda^{-1}$, the effective
supersymmetric theory is parametrized by the squark VEVs, but it
must respect the symmetries of (\ref{glosym}). The symmetries of
the effective interaction is given by the nonperturbative
instanton effects whose symmetry is coming basically from 't Hooft
determinental interaction \cite{tHooft}. The instanton amplitude
is proportional to $e^{-8\pi^2/g^2(\lambda)}\simeq
(\Lambda/\lambda)^{(3N_c-N_f)}$. Here, $\Lambda$ is interpreted as
obtained by integrating the $N_f$ pairs of fermion zero modes.
Therefore, the U(1)$_A$ charge of $\Lambda^{(3N_c-N_f)}$ is
interpreted as that of $N_f$ pairs of squarks. One must have an
appropriate power so that the U(1)$_R$ symmetry is preserved.
Thus,   from the consideration of supersymmetry and  global
symmetries one obtains the effective superpotential as
\cite{DDS83},
\begin{equation}
W_{\rm eff}= C_{N_c,N_f}\left( \frac{\Lambda^{3N_c-N_f}}{{\rm
det.} Q\bar Q}\right)^{1/(N_c-N_f)} \label{Wasymfree}
\end{equation}
where $C_{N_c,N_f}$ is a constant. Eq. (\ref{Wasymfree}) shows the
runaway behavior of the squark fields at low energy for $N_f<N_c$.

Discussions on $N_f\le 3N_c$ are summarized in \cite{IntSeib}.

\subsubsection{Case $N_f>3N_c$}\label{subsub}
On the other hand, if $N_f>3N_c$ then it is asymptotically strong
and the  superpotential given in (\ref{Wasymfree}) does not make
sense as an effective theory. At a larger separation, squarks
behave more freely and the condensation of squarks is not
anticipated. They behave more like free squarks. It is known that
the superpotential given above does not make sense. But the
superpotential we wrote respects all the global symmetries.
Suppose however that we interpret it as the effective
superpotential of free squarks, generated by small-instantons.
Then there results an inconsistency as shown below.

The scale of the small instanton is determined by the coupling
strength at that small-instanton size. For supersymmetric
nonabelian gauge theories, the one loop corrected coupling evolves
as
\begin{equation}
\alpha(\Lambda)=\frac{\alpha}{1+\frac{\alpha}{4\pi}(3N_c-N_f)
\ln(\Lambda^2/\lambda^2)}
\end{equation}
where $\alpha=\alpha(\lambda^2)$. Thus, the instanton amplitude at
the scale $\Lambda^2$ is estimated as
\begin{align}
e^{-8\pi^2/g^2(\Lambda^2)}=e^{-8\pi^2/g^2(\lambda^2)}
e^{-\frac12(3N_c-N_f)\ln
(\Lambda^2/\lambda^2)}=e^{-2\pi/\alpha}\left(
\frac{\Lambda}{\lambda}\right)^{N_f-3N_c}.
\end{align}
If we assign the U(1)$_A$ quantum number of $2N_f$ to
$(\Lambda)^{N_f-3N_c}$ as before, we anticipate a superpotential,
respecting the SU$(N_f)_L\times $SU$(N_f)_R\times
$U(1)$_A\times$U(1)$_B$ symmetry as
$$
W\to e^{-2\pi/\alpha}C_{N_f,N_c} \left(
\frac{\Lambda}{\lambda}\right)^{N_f-3N_c}\left(\frac{1}{{\rm
det.}Q\bar Q}\right) \to
e^{-2\pi/\alpha}\left(\frac{\Lambda^{N_f-3N_c}}{{\rm det.}Q\bar
Q}\right)^{\rm a \ power}
$$
Considering the U(1)$_R$ symmetry also,
 we expect
\begin{equation}
W= e^{-2\pi/\alpha}\lambda^{\frac{4N_f}{N_f-N_c}}\left(\frac{
\Lambda^{N_f-3N_c}}{{\rm det.}Q\bar Q}\right)^{1/(N_f-N_c)}
\end{equation}
However, the vacuum does not exist with
SU($N_f$)$\times$U(1)$_A\times$U(1)$_B \times$U(1)$_R$ because
det.$\ Q\bar Q=0$.  This arose by imposing supersymmetry and
matching the global charges of the original and the effective
theories. So the supersymmetry with the meson condensation is
inconsistent with the asymptotically strong in the ultraviolet or
asymptotically free in the infrared region. In fact, it was too
far fetched to introduce a nonperturbatively generated VEVs for
squark condensates where the theory is infrared-free.

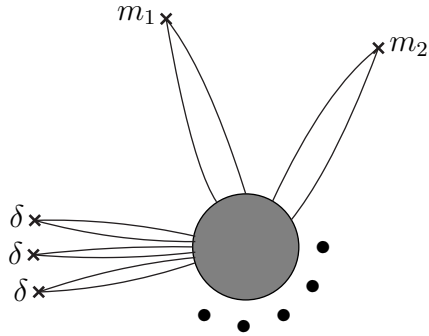
\begin{figure}[t]
\begin{center}
\begin{picture}(400,173)(0,0)

\GCirc(180,60){20}{0.5} \Curve{(150,146)(155,120)(169,77)}
\Text(147,148)[r]{$m_1$} \Curve{(150,146)(175,95)(180,80)}
  \Curve{(230,135)(210,113)(190,77)}
  \Curve{(230,135)(210,90)(197,70)}
\Text(235,136)[l]{$m_2$}

\Curve{(100.5,70)(130,69)(161,64)}
\Curve{(100.5,70)(130,64)(161,62)}
\Curve{(100,57)(130,59.5)(160,60)}
\Curve{(100,57)(130,56)(161,58)}
 \Curve{(102,43)(130,51)(161,56)}
\Curve{(102,43)(130,46)(161,54)}

\Text(97,72)[r]{$\delta$}\Text(96.5,58)[r]{$\delta$}
\Text(98.5,44)[r]{$\delta$}

\SetWidth{1} \Line(152,148)(148,144)\Line(152,144)(148,148)
\Line(232,133)(228,137)\Line(232,137)(228,133)

\Line(102.5,72)(98.5,68)\Line(102.5,68)(98.5,72)
\Line(102,59)(98,55)\Line(102,55)(98,59) \Line(104,45)(100,41)
\Line(104,41)(100,45)

\SetWidth{0.5} \Text(165,34)[c]{$\bullet$}
\Text(180,30)[c]{$\bullet$} \Text(195,34)[c]{$\bullet$}
\Text(206,45)[c]{$\bullet$} \Text(210,60)[c]{$\bullet$}

\end{picture}
\caption{Closing the fermion loops by current masses, leading to
the power $m^{N_f}m_{\tilde G}^{N_c}$.}\label{CurrMass}
\end{center}
\end{figure}

Now, we may include the 't Hooft vertex directly, and introduces
soft supersymmetry breaking with mass splitting of order
$\delta^2$.
\begin{figure}[h]
\begin{center}
\begin{picture}(400,173)(0,0)

\GCirc(180,60){20}{0.5} \ArrowLine(122,140)(169,77)
\Text(120,140)[r]{$Q_1$}

\ArrowLine(178,157)(180,80) \Text(179,168)[c]{$\bar Q_1$}
  \ArrowLine(238,140)(190,77)
\Text(241,140)[l]{$Q_2$}

\ArrowLine(100,60)(160,60) \ArrowLine(100.5,67)(161,62)
\ArrowLine(100.5,53)(161,58) \ArrowLine(101,46)(161,56)
\ArrowLine(101,74)(161,64)

\Text(94,60)[r]{$\tilde G$} \Text(154,45)[c]{$\bullet$}
\Text(165,34)[c]{$\bullet$} \Text(180,30)[c]{$\bullet$}
\Text(195,34)[c]{$\bullet$} \Text(206,45)[c]{$\bullet$}
\Text(210,60)[c]{$\bullet$}
\Text(206,75)[c]{$\bullet$} 

\end{picture}
\caption{Instanton interaction. There are $2N_f$ quark lines and
$2N_c$ gluino lines.}\label{Instanton}
\end{center}
\end{figure}
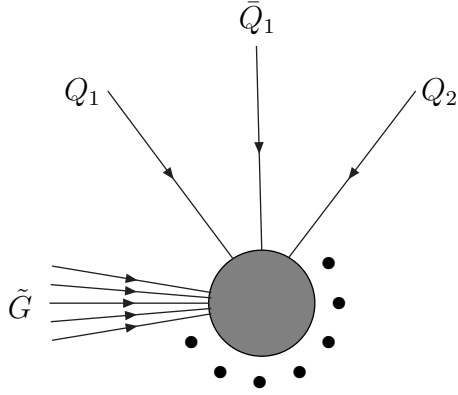
The 't Hooft vertex is shown in Fig. \ref{Instanton}. The effect
of the short distance instanton interaction is obtained by closing
the quark loops and gluino loops with the insertion of their
current masses, which is shown in Fig. \ref{CurrMass}. It gives a
power of masses of the form
$$
\propto\ m_{\tilde G}^{N_c}\cdot {\rm det.}M_{Q}.
$$
where $m_{\tilde G}$ is the gluino mass scale. However, we must
pick up a term of order $\delta^2$, taking into account of the
soft supersymmetry breaking. Diagonalizing the quark mass matrix,
we have the contribution to the vacuum energy as\footnote{The
quark masses are considered to be small compared to the inverse
size, $\rho^{-1}$, of the instanton.}

\begin{equation}
V(\rho)\simeq\lambda^{4-N_f-N_c}  m_{\tilde G}^{N_c}|m_1m_2\cdots
m_{N_f}|\left(1-\cos(\theta_{\tilde
G}+\theta_1+\theta_2+\cdots+\theta_{N_f})\right) \label{DetMInt}
\end{equation}
where we  fine-tuned the vacuum energy so that the SU($N_c$)
vacuum angle $\sum_i\theta_i=0$ corresponds to the minimum. Thus,
we can take the effective interaction by integrating out with the
instanton size from the string scale $M_s^{-1}$ to $m_q^{-1}$,
\begin{equation}
V_{\rm eff}\simeq\int_{1/M_s}^{1/m_q}\frac{d\rho}{\rho^5}
D(\rho)\lambda^{4-N_f-N_c}  m_{\tilde G}^{N_c}|m_1m_2\cdots
m_{N_f}|\left(1-\cos(\theta_{\tilde
G}+\theta_1+\theta_2+\cdots+\theta_{N_f})\right) \label{Deteff}
\end{equation}
where $D(\rho)$ is the density factor of the small-scale
instantons. Taking $D(\rho)$ as \cite{CDGdil}
\begin{equation}
D(\rho)=C_N\left(\frac{8\pi^2}{g^2(\rho^{-1})}\right)^{2N_c}\ ,\ \
[C]=(\rm mass)^{-4}
\end{equation}
for SU($N$),
\begin{align}
V_{\rm dilute}\simeq
&\int_{1/M_s}^{1/m_q}\frac{d\rho}{\rho^5}D(\rho)e^{-2\pi/\alpha(\lambda)}
\lambda^{4-N_f-N_c}  m_{\tilde G}^{N_c}|m_1m_2\cdots
m_{N_f}|\left(1-\cos\theta\right)\\
\to &e^{-2\pi/\alpha(\lambda)}
\frac{C_N}{4}\left(\frac{2\pi}{\alpha(1/
\bar\rho)}\right)^{2N_c}(M_s^4-m_q^4) \lambda^{4-N_f-N_c}
m_{\tilde G}^{N_c}|m_1m_2\cdots m_{N_f}|\left(\cdots\right)
\label{dilute}
\end{align}
where $\theta=\theta_{\tilde
G}+\theta_1+\theta_2+\cdots+\theta_{N_f}, \bar\rho$ is an
appropriate average scale and in the last line we neglected the
$\rho$ dependence of $\lambda$. Note that $V_{\rm dilute}$ is
negligible for SU(2)$_W$ and SU(3)$_c$ gauge groups since there
exist light leptons and light quarks. For this mechanism to be
useful at all, there should be an {\it additional} nonabelian
group at high energy scale with its spectrum vectorlike. For this
to be applicable to Higgsino pairs, Higgsinos must carry the
vectorlike quantum numbers under this nonablian gauge group. This
must be broken above the GUT scale \MGUT. So, we may take
$m_{\tilde G}\sim $\MGUT. Hence the small instanton solution in
the additional nonabelian group does not extends to infinity as in
the case of unbroken nonabelian groups but extends only up to
$M_{\rm GUT}^{-1}$. Nevertheless, the profile of the instanton
solution of the additional nonabelian group for a scale $\ll
M_{\rm GUT}^{-1}$ is very similar to that of the unbroken gauge
group, which is understood below.

The determinental interaction $|m_1m_2\cdots m_{N_f}|$ in
(\ref{dilute}) takes a minimum when at least one  quark mass
vanishes.\footnote{Here, `quark' corresponds to Higgsino.} Thus,
we obtain degenerate vacua, corresponding to
\begin{align}
{\rm Case\ 1}:\ &m_1 =0,\ m_2=m_3=\cdots\ne 0\label{MSSM}\\
{\rm Case\ 2}:\ &m_1=m_2=0,\ m_3=\cdots\ne 0,\ \\
&\rm etc.\nonumber
\end{align}
The vacuum of Case 1 chooses one pair of Higgs doublets by
cosmologically sliding down the massion field $s$, and the vacuum
of Case 2 chooses two pairs of Higgs doublets, etc. Certainly,
Case 1 belongs to the acceptable vacuum, leading to one pair of
Higgs doublets.

The expression (\ref{dilute}) is dominated by the smallest size
instantons since the density of smaller size instantons is larger
than the larger size instantons. If we included the $\rho$
dependence of $\lambda$ in the estimation of (\ref{dilute}), the
importance of the small size instantons in asymptotically strong
theories is more conspicuous.

If we take $\alpha=1/25$ which is the value obtained at the scale
$M_{\rm GUT}$ by extending the low energy couplings in SUSY GUTs,
the GUT scale instantons would contribute as $10^{-50}M_{\rm
GUT}^4$ which is utterly negligible compared to the supergravity
parameter $M_P^2M_W^2$. But, for an illustration, consider the
case of $e^{-2\pi/\alpha}\sim \frac{1}{10}$ at $\rho^{-1}=\frac12
M_s$, for which $\alpha=2.73$. Since loop corrections are expected
to appear as powers of $\alpha/2\pi$, this value can be considered
as the boundary value for a perturbative calculation. Then,
setting every unknown mass parameter in Eq. (\ref{dilute}) as
$\frac12 M_s$, Eq. (\ref{dilute}) gives the height of the
potential as
$$
\sim \frac{32\pi^4 }{5}\delta^{N_c} M_s^{4-N_c}(4\pi)^{2(N_c-2)}.
$$
If $N_c\ge 3$, the contribution of the small size instantons is
negligible.  On the other hand, if $N_c=2$, the instanton
contribution can dominate, by a factor of $10^5-10^6$, the one
loop contribution of Sec. \ref{Sec:OneLoop} due to the
 $1/64\pi^2$ factor present in (\ref{exponeloop}) and a large
numerical factor in Eq. (\ref{dilute}). It is the dynamical
realization of the doublet-triplet splitting, which was first put
forward as an ansatz in Ref. \cite{cchk}. If Case (i) of
(\ref{MSSM}) is chosen, then the MSSM results. Of course, this
conclusion depends on the assumption of the strong gauge couplings
at the string scale.

But the above type small-size instantons involving the SM quarks
and leptons are completely negligible  because the SM quark and
lepton masses are less than TeV. One must employ another
nonabelian gauge group which is broken at or above \MGUT.

\subsubsection{A supergravity toy model}

Let us consider an SU(2)$_R\times$SU(2)$_L\times $U(1)$_{Y_R}$
gauge theory with the following representations,
\begin{align}
\H=({\bf 2,2})\ &:\ N_H {\ \rm flavors\ with\ }Y_R=0\\
  R^c = ({\bf 2,1})\ &:\ N_R {\ \rm flavors\ with\ }Y_R=+\frac12\\
 R = ({\bf 2,1})\ &:\ N_R {\ \rm flavors\ with\ }Y_R=-\frac12\\
 l=({\bf 1,2})\ &:\ N_g {\ \rm flavors\ with\ }Y_R=-\frac12 \\
e^c = ({\bf 1,1})\ &:\ N_g {\ \rm flavors\ with\ }Y_R=+1,\\
{\rm an}&{\rm d} \ N_g {\ \rm flavors\ of\ quarks}
\end{align}
where $N_g= 3$. The electromagnetic charge is
$$
Q_{\rm em}=T_{R3}+T_{L3}+Y_R.
$$
The SU(2)$_R$ symmetry is broken at \MGUT\ with an $\langle
R\rangle\sim $\MGUT. $(R^c+R)$ is vectorlike and can be removed at
 \MGUT.  At low energy, we have the SM gauge group with the usual
$N_g$ lepton families. The small-scale SU(2)$_L$ instanton
interaction is negligible due to the lightness of quark and lepton
masses. But the small-scale SU(2)$_R$ instantons would emit $\H$,
$R^c$ and $R$. Naively, one would expect
\begin{equation}
\propto m^2_{\tilde G_R} {\rm det.}M_{\tilde\H}\label{SUGRAex}
\end{equation}
where $m^2_{\tilde G_R}$ is of order \MGUT\ since SU(2)$_R$ is
broken at the GUT scale. For the SU(2)$_R$ breaking, we need VEVs
of $R$ and $R^c$  which are expected to be heavy at \MGUT. This
kind of mass insertions are understood in Eq. (\ref{SUGRAex}).
However, it should be further suppressed since supersymmetry
restricts the SUSY mass splitting $\delta^2$ appear in the
potential. Thus, we will obtain instead
\begin{equation}
\propto \delta^2 {\rm det.}M_{\tilde\H}.
\end{equation}
Then following the previous argument, we obtain at least one light
Higgs doublet. A nonrenormalizable superpotential of the following
form
$$
\frac{\langle R\rangle}{M_s}\H (le^c\ {\rm and}\ qd^c)\ ,\
\frac{\langle R^c\rangle}{M_s}\H ( qu^c)
$$
can give  mass to the SM fermions where $\H$ is the light Higgs
doublet pair.

\subsubsection{TeV scale Higgsino mass}

Case (i) of (\ref{MSSM}) for $N_c=2$ realizes one light Higgsino
pair.\footnote{Here, $N_c$ corresponds to the nonabelian group
SU($N_c$).} The color triplet Higgsinos are made superheavy not
affected by the determinental interaction. But, for this scenario
to be made successful, the one loop contribution will not spoil
the condition of $m_1\simeq 0$. If the potential of massion $s$
from the determinental interaction is contaminated by other terms,
one must ensure that the other terms do not spoil in choosing one
massless pair of Higgs doublets. Suppose that they are composed of
two terms, the determinental one from the small instanton
contribution and the other from the one-loop contribution. Let us
parametrize them by cosine potentials as
$$
V=-A\left[\cos\left(\frac{s}{M}\right)+\epsilon
\cos\left(\frac{s}{M}-\eta\right) \right]
$$
where $\eta$ is a mismatch phase between the two terms, and
$\epsilon$ is expected to be of order $10^{-6}$. With
$\epsilon=0$, one obtains $s=0$ which corresponds to a zero
Higgsino mass. For a small $\epsilon$, the minimum of the
potential occurs at $s/M\simeq\epsilon\eta.$ For $\eta\le 10^{-6}
$, the Higgsino mass is less than $\sim 10^{-12}M\sim 100$ TeV. So
an alignment of $\eta$ close to 0 is needed to achieve a
reasonable Higgsino mass. If only one massion couples to the
massless Higgsino pair but not to the other pairs, then these two
potentials are aligned, which is the case for the example
discussed in Sec. \ref{Sec:tri}.

\section{Instantons not  emitting quarks and leptons}\label{Sec:tangled}

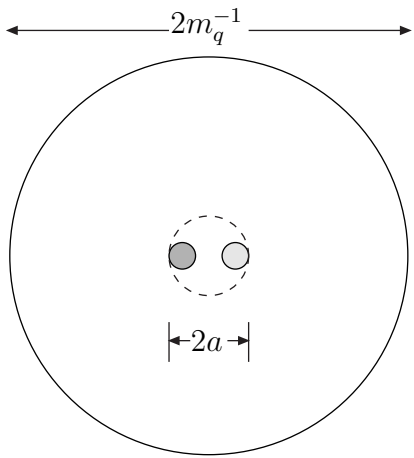
\begin{figure}[h]
\begin{center}
\begin{picture}(400,175)(0,0)
\GCirc(200,75){75}{1}
\LongArrow(185,160)(125,160)\LongArrow(215,160)(275,160)
\Text(200,162)[c]{$2m_q^{-1}$}

\Line(185,50)(185,35)\Line(215,50)(215,35) \Text(200,43)[c]{$2a$}
\LongArrow(193,43)(186,43)\LongArrow(207,43)(214,43)

\GCirc(190,75){5}{0.7}\GCirc(210,75){5}{0.9}
\DashCArc(200,75)(15,0,360){3}

\end{picture}
\caption{Two different instantons (darker grey and lighter grey
ones) separated by $a\gg\rho_{R,L}$ within the space
$\lambda^{-1}=m_q^{-1}$. A dashed boundary with a radius $\simeq
a$ describes an effective size for the tangled
instanton.}\label{sepinstanton}
\end{center}
\end{figure}

The instanton solution is a mapping from the group space $S_3$ to
an Euclidian space-time $S_3$. If we have two instantons one that
of SU(2)$_R$ and the other that of SU(2)$_L$, both of them are
good instanton solutions. These `two' instantons carry both SU(2)
group indices, but their centers can be different. Suppose that
this composite instanton emits $\tilde\H$. If these instantons sit
on top of each other, then the larger size instanton roughly sets
the scale for emitting $\tilde\H$. If the distance of their
separation is larger than the instanton sizes, then the separation
distance roughly sets the scale for emitting $\tilde\H$, which is
schematically shown in Fig. \ref{sepinstanton}.
These composite instantons have three sizes, two instanton scales
$\rho_{R,L}$ and the distance $a$ between them. The above
composite instanton is helpful in introducing instantons with
semi-simple groups.

Our interest is to find out some instantons by which $\tilde\H$ is
emitted but quarks and leptons are not. Since the Higgsino in the
trinification model and in the Pati-Salam model transforms as
$(\bf 2,2)$ under SU(2)$_R\times$ SU(2)$_L$ while quarks and
leptons do not transform in that way, we must utilize this $(\bf
2,2)$ property of $\tilde\H$.

The $S_3$ manifold in the group space is possible with SU(2). For
an instanton solution embedding, let us call the relevant group
SU(2)$_{\rm inst}$. SU(2)$_{\rm inst}$ can be embedded in
SU(2)$_R\times$ SU(2)$_L$ either by identifying ${\bf
2}_R\rightarrow {\bf 2}_{\rm inst}$ and ${\bf 2}_L\rightarrow {\bf
2}_{\rm inst}$, or by identifying ${\bf 2}_R\rightarrow {\bf
2}_{\rm inst}$ and ${\bf 2}_L^*\rightarrow {\bf 2}_{\rm inst}$,
i.e. by identifying SU(2)$_R$ with SU(2)$_L$ or with
SU(2)$_L^*$.\footnote{We encounter this kind of indentification in
the spontaneosly broken SU(2)$_R\times$SU(2)$_L$ by the linkage
Higgs fields $\langle{\bf (2,2^*)} \rangle$ or $\langle{\bf (2,2)}
\rangle$. For instanton solutions, however, we do not need these
Higgs fields.} Let us call this process of identification
`tangling' and the resulting instanton a `tangled instanton'. A
tangled instanton has three sizes, two instanton scales
$\rho_{R,L}$ and the distance $a$ between them. The largest among
these is roughly the effective size of the tangled instanton,
$\rho_t$. We can use $\rho_t$ for the instanton amplitude we
discussed in Subsec. \ref{Subsec:Actsmall}. For the gauge
coupling, we  adopt the most simple choice below: $g_R=g_L$ at the
instanton scale. From now on, we do not use the concept of
 composite instantons. Just, the group property of SU(2)$_{\rm
 inst}$ is important.

The Pontryagin number $q=1$ of SU(2)$_{\rm inst}$ corresponds to
the original Pontryagin number 2 instanton since the gauge
coupling of the diagonal subgroup is reduced by the factor
$\frac{1}{\sqrt2}$ and hence the tunneling amplitude
$\exp[-\frac{8\pi^2}{g^2}q]$ is the same in both interpretations.

Originally, $N_R$ quarks $\psi_{R,R^c}$ of SU(2)$_R$, $N_L$ quarks
$\psi_{L,L^c}$ of SU(2)$_L$, and  $N_H$ Higgsino pairs $\tilde\H$
have the following flavor symmetry
\begin{equation}
SU(N_R)\times SU(N_R)\times SU(N_L)\times SU(N_L)\times
SU(N_H)\label{GlobalS}
\end{equation}
where $U(1)$s are not written.
\begin{figure}[h]
\begin{center}
\begin{picture}(400,120)(0,0)
 \GCirc(200,60){10}{0.5}

\Line(197,80)(197,69) \Line(200,80)(200,70) \Line(203,80)(203,69)
\Line(197,83)(197,95) \Line(200,83)(200,95) \Line(203,83)(203,95)
 \Text(200,105)[c]{SU(2)$_R$} \Line(160,66)(192,66)
\Line(160,63)(191,63) \Line(160,60)(190,60) \Line(160,57)(191,57)
\Line(160,54)(192,54) \Text(157,67)[r]{$\psi_{L,L^c}$}
\Text(157,55)[r]{$\psi_{R,R^c}$} \Line(209,63)(240,63)
\Line(210,60)(240,60) \Line(209,57)(240,57)
\Text(245,62)[l]{$\tilde\H$}
 \Line(197,40)(197,51) \Line(200,40)(200,50)
\Line(203,40)(203,51) \Line(197,37)(197,25) \Line(200,37)(200,25)
\Line(203,37)(203,25) \Text(200,10)[c]{SU(2)$_L$}

\end{picture}
\caption{Tangled instanton with Pontryagin number 1. Solid lines
correspond to quarks, and broken lines correspond to gluinos.
}\label{Pont1}
\end{center}
\end{figure}
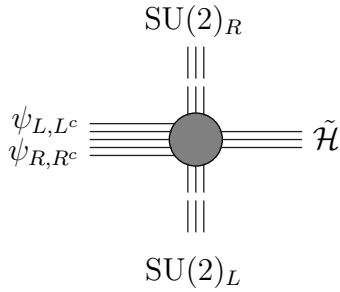
 The SU(2)$_{\rm inst}$ instantons with
Pontryagin number 1 must satisfy the global symmetry
$$
SU(N_R+N_L+N_H)\times SU(N_R+N_L+N_H).
$$
Including gluinos, it is schematically shown in Fig. \ref{Pont1}.
Thus, we expect an instanton generated determinental interaction
from (instanton)$_R\times$(instanton)$_L$ after integrating out
the fermion lines as, identifying SU(2)$_R$ with
SU(2)$_L$,\footnote{It is straightforward to obtain determinental
interactions for the case of identifying SU(2)$_R$ with
SU(2)$_L^*$, which is not discussed here explicitly.}
\begin{equation}
\propto e^{i(\theta_R+\theta_L)}{\rm det.} m_{\tilde G_R}\times
{\rm det.} m_{\tilde G_L}\times{\rm det.} m_{R}\times {\rm det.}
m_{L}\times{\rm det.} m_{\tilde \H}+{\rm h.c.}
\end{equation}
where h.c. is from $\overline{\rm
(instanton)}_R\times\overline{\rm (instanton)}_L$. Masses are
those of SU(2)$_R$ and SU(2)$_L$ gauginos, SU(2)$_R$ and SU(2)$_L$
quarks, and Higgsinos. From the tangled ${\rm
(instanton)}_R\times\overline{\rm (instanton)}_L$ and its tangled
anti-instanton, we would have
\begin{equation}
\propto e^{i(\theta_R-\theta_L)}{\rm det.} m_{\tilde G_R}\times
{\rm det.} m_{\tilde G_L}^*\times{\rm det.} m_{R}\times {\rm det.}
m_{L}^* +{\rm h.c.}
\end{equation}
which does not contain ${\rm det.} m_{\tilde \H}$ because a chiral
transformation of $\tilde\H$ rotates only a combination
$\theta_R+\theta_L$. Certainly, these forms are consistent with
the original global symmetry by assigning appropriate global
transformation properties on the mass matrices.  So far we
considered the SU(2)$_{\rm inst}$ instanton with Pontryagin number
1. The Pontryagin number 1 instantons emit doublets of SU(2)$_{\rm
inst}$.

How about the SU(2)$_{\rm inst}$ instantons with  Pontryagin
number greater than 1? The Pontryagin number 2  SU(2)$_{\rm inst}$
instatons are just two Pontryagin number 1 instantons which is a
trivial extension of Pontryagin number 1 instantons.

The next simple representation of SU(2)$_R$(similarly for
SU(2)$_L$) is the $3\times 3$ representation which gives the
Pontryagin number 4 instanton.\footnote{The SU(2) embedding in
SU(3) was considered in  \cite{Wilczek} where the Pontryagin
number 4 instanton is constructed by four Pontryagin number 1
instantons. Here, we consider SU(2) and such a composition is not
possible.} The Pontryagin index $q$ is given by
\begin{equation}
q=\frac{1}{8\pi^2}{\rm Tr}\int F_{\mu\nu}\tilde F_{\mu\nu} d^4x
\end{equation}
where $\tilde F_{\mu\nu}=\frac12
\epsilon_{\mu\nu\rho\sigma}F_{\rho\sigma}$. The well-known
$2\times 2$ representation of  SU(2)$_{\rm inst}$ gives $q=\pm 1$
with
\begin{equation}
A_\mu=\frac{x^2}{x^2+\lambda^2}g^{-1}\partial_\mu g,\ \ g=
\frac{x_4-i{x_i\sigma_i}}{r}=- \frac{i}{r}x_\mu\sigma_\mu,\
g^{-1}=\frac{i}{r}x_\mu\sigma_\mu
\end{equation}
where $\sigma^i$ are the ordinary Pauli matrices and the SU(2)
generators are $T_i=\frac12\sigma_i.$ For a general SU(2)
representation $T_i$,
\begin{equation}
F_{\mu\nu}=\frac{4\lambda^2}{x^2+\lambda^2}T_{\mu\nu}
\end{equation}
where
\begin{equation}
T_{ij}={i}[T_i,T_j]\ ,\ \ T_{i4}=-T_{4i}=-T_i
\end{equation}
whence for a self-dual field $F_{\mu\nu}=\tilde F_{\mu\nu}$,
\begin{equation}
q=\frac{1}{6}\ {\rm Tr}\ T_{\mu\nu}T_{\mu\nu}.\label{qTexp}
\end{equation}
For the doublet representation  $T_i=\frac12\sigma_i,$ we obtain
$q=1$. For a triplet representation, $q=4$.

In the tangling process of ${\bf 3}_R$ and ${\bf 3}_L$ instantons,
i.e. by identifying SU(2)$_R$ and SU(2)$_L$, we have  two
Pontryagin number 4 intstantons in the original groups, i.e. the
total Pontryagin number 8. However, in terms of SU(2)$_{\rm inst}$
it is a Pontryagin number 4 instantion. As commented before, the
tunneling amplitude is the same whichever interpretation we use
since the SU(2)$_{\rm inst}$ gauge coupling is smaller by a factor
of $\frac{1}{\sqrt2}$. In this case, the triplets of SU(2)$_R$ and
SU(2)$_L$ corresponding to the gluinos transform as a triplet of
SU(2)$_{\rm inst}$,
\begin{align}
{\bf 3}_R\ {\rm of}\ SU(2)_R &\longrightarrow {\bf 3}_{\rm inst}\
{\rm of}\
SU(2)_{\rm inst}\label{Rgluino}\\
{\bf 3}_L\ {\rm of}\ SU(2)_L &\longrightarrow {\bf 3}_{\rm inst}\
{\rm of}\ SU(2)_{\rm inst}\label{Lgluino}
\end{align}
The forms(instanton configuraion) of the triplet gauge field do
not couple to a doublet(spinor) of SO(3) since spinors cannot be
obtained by tensor products of the vector {\bf 3}, i.e. we cannot
write
$$
(i\dsl+\gamma^\mu A_\mu)\Psi=0
$$
for a $3\times 3$ matrix $(A_\mu)^i_j\ (i,j=1,2,3)$ and
$\Psi_\alpha\ (\alpha=1,2)$. On the other hand, if $\Psi$ is
represented as a tensor product of the vector {\bf 3}, for example
$\Psi_{ij}$, then the above type of Dirac equation can be written.
Thus, the Pontryagin number 4 instantons(${\bf 3}_{\rm inst}$) do
not emit $({\bf 2,1})$ and $({\bf 1, 2})$. In our case, the
Higgsinos behave differently from the quark and leptons. They
transform as a {\bf 3},
\begin{align}
 {\bf 2}_R\times{\bf
2}_L  &\longrightarrow {\bf 3}_{\rm inst}+{\bf 1}\ {\rm of}\
SU(2)_{\rm inst}\label{RLHiggsino}
\end{align}

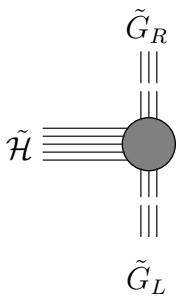
\begin{figure}[h]
\begin{center}
\begin{picture}(400,120)(0,0)
 \GCirc(200,60){10}{0.5}

\Line(197,80)(197,69) \Line(200,80)(200,70) \Line(203,80)(203,69)
\Line(197,83)(197,95) \Line(200,83)(200,95) \Line(203,83)(203,95)
 \Text(200,105)[c]{$\tilde G_R$} \Line(160,66)(192,66)
\Line(160,63)(191,63) \Line(160,60)(190,60) \Line(160,57)(191,57)
\Line(160,54)(192,54) \Text(157,60)[r]{$\tilde\H$}

 \Line(197,40)(197,51) \Line(200,40)(200,50)
\Line(203,40)(203,51) \Line(197,37)(197,25) \Line(200,37)(200,25)
\Line(203,37)(203,25) \Text(200,10)[c]{$\tilde G_L$}

\end{picture}
\caption{Tangled instanton with Pontryagin number 2. Solid lines
correspond to quarks, and broken lines correspond to gluinos. The
emitted quarks are only Higgsinos.}\label{Pont2}
\end{center}
\end{figure}

Thus, in view of (\ref{Rgluino}) and  (\ref{Lgluino}), SU(2)$_R$
and SU(2)$_L$ gluinos are emitted by SU(2)$_{\rm inst}$
instantons, and in view of (\ref{RLHiggsino}) $\tilde\H$ are
emitted by SU(2)$_{\rm inst}$ instantons. However, ${\bf
2}_R=\psi_{R},\psi_{R^c}$ and ${\bf 2}_L=\psi_{L},\psi_{L^c}$ are
not emitted by the SU(2)$_{\rm inst}$ instantons with Pontryagin
number 4.  Thus, the interaction we obtain from tangled instantons
with Pontryagin number 4 with the identification of SU(2)$_R$ with
SU(2)$_L$ is
\begin{equation}
\propto  e^{i(\theta_R+\theta_L)}{\rm det.} m_{\tilde G_R}\times
{\rm det.} m_{\tilde G_L}\times{\rm det.} m_{\tilde H}+{\rm h.c.}
\label{Pont2Int}
\end{equation}
which is schematically shown in Fig. \ref{Pont2}.
On the other hand, the interaction from tangled instantons with
Pontryagin number 4 with the identification of SU(2)$_R$ with
SU(2)$_L^*$ would be
$$
\propto  e^{i(\theta_R-\theta_L)}{\rm det.} m_{\tilde G_R}\times
{\rm det.} m_{\tilde G_L}^*+{\rm h.c.}
$$
where ${\rm det.} m_{\tilde H}$ is absent since a chiral rotation
of $\tilde\H$ rotates only the combination of $\theta_R+\theta_L$.

Trinification models and Pati-Salam models give negligible
contributions from  Fig. \ref{Pont1} due to the highly-chiral
nature of the SM quarks and leptons. Thus, among those involving
$\tilde\H$, Fig. \ref{Pont2} is the dominant one. The small-scale
instanton interaction we discussed in previous sections is
attributed to the one coming from Fig. \ref{Pont2}. Note that the
eventual breaking of SU(2)$_R$ at \MGUT\ does not change our
argument since SU(2)$_R\times$SU(2)$_L$ is not broken at $q$ where
$ M_{\rm GUT}< q<M_{s}$ and the size of small instantons we are
considering is roughly $ M_{s}^{-1}\ll M_{\rm GUT}^{-1}$.

In spontaneously broken supersymmetric models, the dominant
instanton contribution will be at most of order $\delta^2{\rm
det.}M$. Indeed, Fig. \ref{Pont2}  gives this order, since the
SU(2)$_L$ gaugino mass contraction is of order $\delta^2$ since
SU(2)$_L$ is broken at the electroweak scale and SU(2)$_R$ gaugino
mass contraction is of order \MGUT\ since SU(2)$_R$ is broken at
the GUT scale by $\langle {(\bf 2,1)}_{\pm 1/2}\rangle$. However,
${(\bf 2,1)}_{\pm 1/2}$ is not emitted by the $q=4$ instantons
since it is a spinor of SO(3)$_{\rm inst}$. In any case whether
${(\bf 2,1)}_{\pm 1/2}$ is emitted(as in the example of Subsec.
\ref{Subsec:Actsmall}) or not(as above), our idea for dynamical
$\mu_H$ works only for this types of SU(2)$_R\times$SU(2)$_L$
theories where SU(2)$_R$ is broken at a high energy scale.
Trinification type models and Pati-Salam type models belong to
this category. But the SU(5)$_{\rm GG}$ and flipped SU(5)
subgroups of E$_6$ cannot be made to work for the dynamical
$\mu_H$ along the line we discussed here.

The tunneling amplitude for the Higgsino mass matrix is
proportional to $e^{-32\pi^2/g_{\rm inst}^2}$ which has to be
significant for the mechanism to work. If the massion has the flat
direction except from this $q=4$ instanton contribution, then this
will settle probably mass of one Higgsino pair near the
electroweak scale.

\section{Example with a trinification model}\label{Sec:tri}
For an illustrative purpose, we adopt a discrete symmetry to
tackle the problem with a reasonable simplicity. In Sec.
\ref{Sec:OneLoop}, an $S_3$ discrete symmetry has been used.
String $Z_3$ orbifold compactification can have indeed this kind
of the $S_3$ symmetry.

\subsection{SU(2)$_R\times$SU(2)$_L$ as a subgroup of SU(3)$^3$}
For an explicit discussion,  from now on let us proceed with the
trinification model of \cite{trikim}. This model has enough
independent directions so that the vacuum can choose a minimum of
the potential. The gauge group is
SU(3)$_1\times$SU(3)$_2\times$SU(3)$_3$ and the quantum numbers of
the spectrum is three times
\begin{equation}
\Psi_{\rm tri}={\bf 27}_{\rm tri}={\bf
(3^*,3,1)+(1,3^*,3)+(3,1,3^*)}\label{trisu3}
\end{equation}
which is denoted as \cite{kimjhep},
\begin{align}
\Psi_l\to \Psi_{(\bar M,l,0)}=&\Psi_{(\bar
1,i,0)}(H_1)_{-1/2}+\Psi_{(\bar 2,i,0)}(H_2)_{+1/2} +\Psi_{(\bar
3,i,0)}(l)_{-1/2}\nonumber\\
&+\Psi_{(\bar 1,3,0)}(N_5)_{0}+\Psi_{(\bar 2,3,0)}(e^+)_{+1}+
\Psi_{(\bar 3,3,0)}(N_{10})_0\label{lhumor}\\
\Psi_q\to \Psi_{(0,\bar l,\alpha)}=&\Psi_{(0,\bar
i,\alpha)}(q)_{+1/6}+
\Psi_{(0,\bar 3,\alpha)}(D)_{-1/3}\label{qhumor}\\
\Psi_a\to
\Psi_{(M,0,\bar\alpha)}=&\Psi_{(1,0,\bar\alpha)}(d^c)_{+1/3}+
\Psi_{(2,0,\bar\alpha)}(u^c)_{-2/3}+\Psi_{(3,0,\bar\alpha)}(\bar
D)_{+1/3}\label{aqhumor}
\end{align}
Here, we also show the standard model fields in brackets. The
running indices are those of three SU(3)s, i.e. $i$ for SU(2)$_W$
and $\alpha$ for SU(3)$_3$. SU(2)$_W$ is the subgroup of the
second SU(3)$_2$ and QCD is the third SU(3)$_3$. The three types
of representations are called humors: lepton-, quark-, and
antiquark-humors with the obvious implication.

Let the trinification group be broken by $\langle
N_{10}\rangle\sim M_s$
to,\footnote{SU(2)$_1\times$SU(2)$_W\times$U(1) is broken down to
the SM gauge group by $\overline{N_5}$ near \MGUT\ \cite{trikim}.}
\begin{equation}
SU(3)_1\times SU(3)_2\times SU(3)_3\to SU(2)_1\times SU(2)_W\times
SU(3)_c\times U(1).
\end{equation}
Here, SU(2)$_1$ is the  SU(2)$_R$ and SU(2)$_W$ is the  SU(2)$_L$
of Sec. \ref{Sec:tangled}. So the term from Fig. \ref{Pont2} of
tangled instantons emits pairs of $\tilde\H_1=\Psi_{(\bar 1,i,0)}$
and $\tilde\H_2=\Psi_{(\bar 2,i,0)}$. The interaction from Fig.
\ref{Pont2} will involve det.$M_{\tilde\H}$.

For the trinification (\ref{trisu3}), the $\mu_H$ terms arise from
the coupling of type $\Psi_l^3$, i.e.
\begin{equation}
N_{10}\H_1\H_2,\label{muH}
\end{equation}
while the triplet $\mu_T$ term appears from the coupling of type
$\Psi_l\Psi_q\Psi_a$, i.e.
\begin{equation}
N_{10}\bar D D.\label{muT}
\end{equation}
Related to SU(2)$_{\rm inst}$, the notable difference of $\H_1$
and $\H_2$ from $D$ and $\overline{D}$ is that the pair $\H_1$ and
$\H_2$ transforms as $({\bf 2,2})$ under the
SU(2)$_1\times$SU(2)$_W$ while $D$ and $\overline{D}$ are
SU(2)$_1$ and SU(2)$_W$ singlets. As commented before all $D$s and
$\overline{D}$s are removed at the GUT scale\footnote{In fact
there is an additional alignment problem with $\H_1$ and $\H_2$,
which will be commented below.} since there is no tangled
instanton interaction involving them only. So, SU(2)$_1$ and
SU(2)$_W$ instantons emit $\tilde \H_1$ and $\tilde \H_2$ pairs
and hence at least one pair of Higgs doublets out of three(or
nine) pairs from three $\Psi_{\rm tri}$s(or ${\bf 27}_{\rm
tri}+{\bf 27}_{\rm tri} +\overline{\bf 27}_{\rm tri}$)
\cite{trikim} can remain light.

However, if the massion couplings to $\H_1\H_2$ and
$D\overline{D}$ are of the same form, then the massion VEV
determined by the small SU(2)$_1\times$SU(2)$_W$ instantons give
the same eigenvalues to $D$ and $\overline{D}$ pairs, and there
will result a light $D$ and $\overline{D}$. But, the origins of
the Yukawa couplings of $\H_1$ and $\H_2$ and the Yukawa couplings
of $D$ and $\overline{D}$ are different, as pointed out in
(\ref{muH}) and (\ref{muT}). So, if an $S_3$ symmetry is imposed,
we expect the couplings of the form given in Eq.
(\ref{MMHiggsino}), but with different sets of $a$ and $b$ of Eq.
(\ref{MMHiggsino}), say $a_H$ and $b_H$  for $\H_1\H_2$ and $a_D$
and $b_D$ for $D\overline{D}$. So, a massless ratio of $a_H$ and
$b_H$ for $\H_1\H_2$ does not necessarily lead to a massless ratio
of $a_D$ and $b_D$ for $D\overline{D}$. Thus, when one Higgsino
pair is made light by a tangled instanton with Pontryagin number
4, there does not necessarily result a light pair of $D$ and
$\overline{D}$. It is a dynamical solution of the MSSM problem.

\subsection{Example with three singlet fields}

Consider just three copies of $\Psi_{\rm tri}$, neglecting $3({\bf
27}_{\rm tri} +\overline{\bf 27}_{\rm tri})$, for the simplicity
of discussion. The singlets generating the $\mu_H$ terms are
$N_{10}$'s of (\ref{lhumor}),
\begin{align}
\sum_{abc}f_{abc}N_{10}^{(a)} H_1^{(b)}H_2^{(c)}
\end{align}
where $a,b,c$ are the flavor indices. Let us choose the Yukawa
couplings  $f_{abc}=f$ for $\H_1\H_2$ so that the calculation is
simple. Namely, we choose $a_H=b_H$, and for this choice we expect
in general $a_D\ne b_D$. The $3\times 3$ Higgsino mass matrix
becomes
\begin{equation}
M_{\tilde H}=\left(
\begin{array}{ccc}
f\langle N_{10}^{(1)}\rangle\ & f\langle N_{10}^{(3)}\rangle\ &
f\langle N_{10}^{(2)}\rangle\\
f\langle N_{10}^{(3)}\rangle \ & f\langle N_{10}^{(2)}\rangle \ &
f\langle N_{10}^{(1)}\rangle \\
f\langle N_{10}^{(2)}\rangle \ & f\langle N_{10}^{(1)}\rangle \ &
f\langle N_{10}^{(3)}\rangle
\end{array}
\right)\label{genHiggsino}
\end{equation}
Let  $v_a\equiv\langle N_{10}^{(a)}\rangle$. Thus, there are three
independent fields and they can settle at the minima near
\begin{align}
{\rm Det.}M_{\tilde
H}=-\frac{f^3}{2}(v_1+v_2+v_3)\left[(v_1-v_2)^2+(v_2-v_3)^2+
(v_3-v_1)^2\right]=0.\label{DetMH}
\end{align}
The eigenvalue $x$ of $M_{\tilde H}$ satisfies
\begin{align}
x^3&-f(v_1+v_2+v_3)x^2+\frac{f^2}{2}[(v_1-v_2)^2+(v_2-v_3)^2
+(v_3-v_1)^2]x\\
&-\frac{f^3}{2}(v_1+v_2+v_3)\left[(v_1-v_2)^2+(v_2-v_3)^2+
(v_3-v_1)^2\right]=0.
\end{align}
 The solutions of Eq.
(\ref{DetMH}) in the $(v_1,v_2,v_3)$ space are
\begin{align}
{\rm (i)} &\ {\rm the\ plane\ }v_1+v_2+v_3=0,\ {\rm except\ the\ origin}\\
{\rm (ii)} &\ {\rm the\ line\ } v_1=v_2=v_3\ ,\ {\rm except\ the\
origin}.
\end{align}
Case (i) gives one pair of massless Higgsinos and Case (ii) gives
two pairs of massless Higgsinos. Therefore, Case (i) allows the
MSSM spectrum with one pair of Higgs doublets at low energy.
Cosmologically, it is more probable for an arbitrary initial set
of $(v_1,v_2,v_3)$ to find the plane configuration before to find
the line configuration. Thus, one light pair of Higgs doublets is
expected to be chosen cosmologically.

Let us discuss Case (i) only, with $v_3=-v_1-v_2$. The mass
eigenstates are
\begin{align}
m_1=&0:\ \psi^{(0)}=\frac{1}{\sqrt3}\left(\begin{array}{c} 1\\
1\\ 1\end{array}\right)\\
m_{\pm}=& \pm
f\sqrt{\frac32\left[v_1^2+v_2^2+(v_1+v_2)^2\right]}\\
&\psi^{(+)}\propto\left(\begin{array}{c} v_1+2v_2\\
-m_++v_1-v_2\\ m_+-2v_1-v_2\end{array}\right)\\
&\psi^{(-)}\propto\left(\begin{array}{c} v_1+2v_2\\
m_++v_1-v_2\\ -m_+-2v_1-v_2\end{array}\right)
\end{align}

For Case (i), let the massions are chosen as
\begin{align}
S_0=&\frac{1}{\sqrt3}\left(N_{10}^{(1)}+N_{10}^{(2)}+N_{10}^{(3)}
\right)\\
S_{+,-}=&{\rm\ orthogonal\ to\ }S_0.
\end{align}
Thus, we may write the mass terms Higgsinos $\psi^{(0)}_i=\tilde
H_i$ as
\begin{align}
\propto\ S_0\psi^{(0)}_1\psi^{(0)}_2+\cdots\label{formY}
\end{align}
where $\cdots$ does not contain $S_0$.

Since $S^{(0)}$ appears only with the massless Higgsino in the
mass matrix, we can study its dependence on mass matrix easily
even we include the one loop correction of (\ref{ScaleLam}).
Adding Eqs. (\ref{Pont2Int}) and (\ref{ScaleLam}), we can pick up
the $S_0$ dependence from the $m^{(0)}$ eigenvalue which is zero
at the \MGUT\ scale, which is parametrized as
\begin{equation}
V(S_0)\propto \delta^2[ m^{(0)}]^2+\gamma \left\{\delta^2[
m^{(0)}]^2+2\delta^2[ m^{(0)}]^2\ln\frac{[
m^{(0)}]^2+\delta^2}{\lambda^2} \right\}\label{IntSum}
\end{equation}
where we set $\delta^2$ in Eqs. (\ref{Pont2Int}) and
(\ref{ScaleLam})  the same. The ratio of the overall interaction
strengths is $\gamma$. Minimization of $V(S_0)$ leads to a nonzero
$ m^{(0)}$ at a value of order $\delta$ possibly corrected by a
logarithmic factor. This is because $V(S_0)$ does not contain any
large number except by a logarithmic factor of $\lambda$.

\subsection{Massion coupling with $S_3$ symmetry}

To show the form (\ref{formY}), it is convenient to use the tensor
product table of $S_3$ \cite{EMa}.\footnote{The permutation
symmetry has been extensively used before in particle physics.
Some references are \cite{PerSymm}.} The $S_3$ representations are
{\bf 1} and {\bf 2}. Thus, three components of Higgsinos can split
into either three {\bf 1}s or a {\bf 1} and a {\bf 2}. The latter
case is of our immediate concern. Let us consider $S_3$ with
elements $\psi^{(a)}$ with $a=1,2,3$. The $S_3$ representations
are
\begin{align}
 {\bf 1}=(1^0)&= \psi^{(0)}=\frac{1}{\sqrt3}
 \left(\psi^{(1)}+\psi^{(2)}+\psi^{(3)}
 \right)\\
 {\bf 2}=\left(\begin{array}{c}
2^+\\ 2^-
 \end{array}\right)&=\left(\begin{array}{c}
\psi^{(+)}\\ \psi^{(-)}
 \end{array}\right)=\frac{1}{\sqrt3}\left(
 \begin{array}{c}
 \psi^{(1)}+\omega\psi^{(2)}+\omega^2\psi^{(3)}\\
 \psi^{(1)}+\omega^2\psi^{(2)}+\omega\psi^{(3)}\\
 \end{array}
 \right).
\end{align}
where $\omega$ is a cube root of unity, $e^{ 2\pi/3}$ (or $e^{
-2\pi/3}$). The tensor product of {\bf 2} is
\begin{equation}
{\bf 2}\times{\bf 2}={\bf 2}+{\bf 1}+{\bf
1}^\prime.\label{Tensorprod}
\end{equation}
The doublet combination of (\ref{Tensorprod}) transforms under an
$S_3$ operation  as
\begin{align}
\left(\begin{array}{c} \psi^{(+)}_1\psi^{(+)}_2\\
\psi^{(-)}_1\psi^{(-)}_2
 \end{array}\right)\longrightarrow\left(\begin{array}{c}
 \omega^2\psi^{(-)}_1\psi^{(-)}_2\\
\omega\psi^{(+)}_1\psi^{(+)}_2
 \end{array}\right)=\left(\begin{array}{cc}
 0&\omega^2\\
\omega &0
 \end{array}\right)\left(\begin{array}{c}
 \psi^{(+)}_1\psi^{(+)}_2\\
\psi^{(-)}_1\psi^{(-)}_2
 \end{array}\right)
\end{align}
where the $2\times 2$ matrix in the last equation is a member of
$S_3$ generators on doublets. On the other hand, the singlet
couplings are
\begin{equation}
2^+\cdot 2^-+2^-\cdot 2^+={\bf 1}\ ,\ \ 2^+\cdot 2^--2^-\cdot
2^+={\bf 1}^\prime\ .\label{Tensor1}
\end{equation}

For the doublet to obtain mass, it must couple to the doublet
components among three $S^{(a)}$:
$S^{(+)}=\frac{1}{\sqrt3}(S^{(1)}+\omega S^{(2)}+\omega^2
S^{(3)})$ or $S^{(-)}=\frac{1}{\sqrt3}(S^{(1)}+\omega^2
S^{(2)}+\omega S^{(3)})$. Thus, inverting this expression for
$a=1,2,3$, we obtain $S^{(a)}=\frac{1}{\sqrt3}(S^{(0)}+\cdots)$,
where $\cdots$ contain only $S^{(+)}$ and $S^{(-)}$. Suppose the
doublet in the RHS of (\ref{Tensorprod}) couples to a doublet of
$S^{(a)}$ to give the combination {\bf 1} of (\ref{Tensor1}).
Then, we obtain
$$
S^{(+)}(\psi^{(+)}_1\psi^{(+)}_2) +
S^{(-)}(\psi^{(-)}_1\psi^{(-)}_2).
$$
This is a proper form for the diagonalized Yukawa couplings. It
should be such that $\langle S^{(\pm)}\rangle\propto m_\pm$.
Namely, $S^{(0)}$ couples only to $\psi^{(0)}_1\psi^{(0)}_2$ as
claimed in Eq. (\ref{formY}) with $\langle S^{(0)}\rangle=0$. The
same conclusion is drawn from the combination {\bf 1}$^\prime$ of
(\ref{Tensor1}).

\section{Conclusion}\label{Sec:conclusion}

We considered the case where gauge couplings become asymptotically
strong near the cutoff scale $M_s$ or $M_P$. The asymptotically
strong gauge coupling can arise due to the presence of a large
number of matter fields below the compactification scale from the
string compactification. In this high energy strong coupling
regime, the small scale instanton contribution to the potential of
the Higgs boson has been considered in this paper. We showed that
the $q=4$ SU(2)$_{\rm inst}$ instanton allows a determinental
interaction of Higgsino mass matrix without the quark and lepton
mass matrices as shown in Fig. \ref{Pont2},
$$
\propto  e^{i(\theta_R+\theta_L)}{\rm det.} m_{\tilde G_R}\times
{\rm det.} m_{\tilde G_L}\times{\rm det.} m_{\tilde H}+{\rm h.c.}
$$
Thus, the minimum of the potential is shown to make some of the
Higgs boson pairs choose mass at zero which would be shifted to
O($\delta^2$) via the soft breaking of supersymmetry. This small
size instantons can be made to work only for the gauge group
SU(2)$_R\times$SU(2)$_L$ which can be a subgroup of the
trinification group SU(3)$^3$ or a subgroup of the Pati-Salam
group SU(2)$_R\times$SU(2)$_L\times$SU(4). In this regard, we
considered a trinification model where the {\it massion} is shown
to couple only to the massless pair of the Higgs doublets. How
many pairs of Higgs doublets are chosen to be light might be
determined from the cosmological consideration. In the
trinification example we considered, the vacuum with one massless
pair is a plane while the vacuum with two massless pairs is a
line, etc., in the $(v_1,v_2,v_3)$ space, and hence cosmologically
it is likely that the plane vacuum is more easily accessible,
making just one massless pair for the Higgs doublets.

\bigskip

\acknowledgments I have benefitted from discussions with Kang-Sin
Choi, Kiwoon Choi, Hyung Do Kim, and Hans Peter Nilles at various
stages of developing the present idea.  This work is supported in
part by the KOSEF Sundo Grant, the KOSEF ABRL Grant No.
R14-2003-012-01001-0, and the BK21 program of Ministry of
Education.


\bigskip

\begin{thebibliography}{999}

\def\apj#1#2#3{Astrophys.\ J.\ {\bf #1} (#3) #2}
\def\ijmp#1#2#3{Int.\ J.\ Mod.\ Phys.\ {\bf #1} (#3) #2}
\def\mpl#1#2#3{Mod.\ Phys.\ Lett.\ {\bf A#1} (#3) #2 }
\def\nat#1#2#3{Nature\ {\bf #1} (#3) #2}
\def\npb#1#2#3{Nucl.\ Phys.\ {\bf B#1} (#3) #2}
\def\plb#1#2#3{Phys.\ Lett.\ {\bf B#1} (#3) #2}
\def\prd#1#2#3{Phys.\ Rev.\ {\bf D#1} (#3) #2}
\def\prl#1#2#3{Phys.\ Rev.\ Lett.\ {\bf #1} (#3) #2}
\def\prt#1#2#3{Phys.\ Rep.\ {\bf #1} (#3) #2}
\def\sjnp#1#2#3{Sov.\ J.\ Nucl.\ Phys.\ {\bf #1} (#3) #2}
\def\zp#1#2#3{Z.\ Phys.\ {\bf C}{\bf #1} (#3) #2}
\def\jhep#1#2#3{JHEP\ {\bf #1} (#3) #2}
\def\epjc#1#2#3{Euro. Phys. J.\ {\bf C#1} (#3) #2}
\def\rmp#1#2#3{Rev. Mod. Phys.\ {\bf #1} (#3) #2}

\bibitem{mu} J. E. Kim and H. P. Nilles, {\it The $\mu$ problem and
the strong CP problem}, \plb{138}{150}{1984};\\
G. Giudice and A.
Masiero, {\it A natural solution to the $\mu$ problem in
supergravity theories}, \plb{206}{480}{1988}.

\bibitem{dimgeo} S. Dimopoulos and H. M. Georgi, {\it Softly broken
supersymmetry and SU(5)}, \npb{193}{150}{1981};\\
N. Sakai, {\it Naturalness in supersymmetric GUTs},
\zp{11}{153}{1981}.

\bibitem{iknq} L. Iba\~nez, J. E. Kim, H. P. Nilles, and F.
Quevedo, {\it Orbifold compactifications with three families of
$SU(3)\times SU(2)\times U(1)^N$}, \plb{191}{282}{1987}.


\bibitem{PQVW} R. D. Peccei and H. R. Quinn, {\it CP conservation in the
presence of instantons},\prl{38}{1440}{1977};\\
C. Vafa and E. Witten, {\it Parity conservation in QCD},
\prl{53}{535}{1984}.\\
For a review, see, J. E. Kim,
\prt{150}{1}{1987}.

\bibitem{cchk} K.-S. Choi, K.-Y. Choi, K. Hwang, and J. E. Kim, {\it
Higgsino mass matrix ansatz for MSSM}, \plb{579}{165}{2004}.

\bibitem{choihd} M. Dine and N. Seiberg, {\it String theory and
the strong CP problem}, \npb{273}{109}{1986}; \\
J. Flynn and L. Randall, {\it A computation of the small instanton
contribution to the axion potential}, \npb{293}{731}{1987};\\
K. Choi and H. D. Kim, {\it Small instanton contribution to the
axion potential in supersymmetric models}, \prd{59}{072001}{1999}.

\bibitem{WZgr} S. Deser and B. Zumino, {\it Broken supersymmetry and
supergravity}, \prl{38}{1433}{1976}.

\bibitem{musym} J. E. Kim and H. P. Nilles, {\it Symmetry principles
toward solutions of the $\mu$ problem}, \mpl{9}{3575}{1994}.

\bibitem{IntSeib} K. A. Intriligator and N. Seiberg,
{\it Lectures on supersymmetric gauge
theories and electric-magnetic duality}, Nucl. Phys. Suppl. {\bf
45BC} (1996) 1.

\bibitem{DDS83} A. C. Davis, M. Dine and N. Seiberg, {\it
The massless limit of supersymmetric QCD}, \plb{125}{487}{1983}.

\bibitem{tHooft} G. 't Hooft, {\it Computation
of quantum effects due to a four-dimensional pseudoparticle},
\prd{14}{3432}{1976}.

\bibitem{CDGdil} C. G. Callan, R. Dashen, and D. J. Gross,
{\it Towards a theory of the strong
interactions}, \prd{17}{2717}{1978}.

\bibitem{Wilczek} F. Wilczek, {\it Inequivalent
embeddings of SU(2) and instanton interactions},
\plb{65}{160}{1976}.

\bibitem{trikim} J. E. Kim, {\it Trinification with
$\sin^2\theta_W=\frac38$
and seesaw neutrino mass}, \plb{591}{119}{2004}.

\bibitem{kimjhep} J. E. Kim, {\it SU(3) trits of orbifolded
$E_8\times E_8^\prime$ heterotic string and supersymmetric
standard model}, \jhep{0308}{010}{2003}.

\bibitem{EMa} For a recent review, see, E. Ma, hep-ph/0409075, {\it
Nonabelian discrete family symmetries of leptons and quarks}.

\bibitem{PerSymm} 
S. Pakvasa and H. Sugawara, {\it Discrete symmetry and Cabibbo
angle}, \plb{73}{61}{1978};\\
G. Segr\`e and H. A. Weldon, {\it Natural suppression of strong P
and T violations and caculable mixing angles in $SU(2)\times
U(1)$}, \prl{42}{1191}{1979};\\
P. F. Harrison and W. G. Scott, {\it Generation permutation
symmetry and the quark mixing matrix}, \plb{333}{471}{1994};
\\
P. F. Harrison, D. H. Perkins and W. G. Scott, {\it Threefold maximal
lepton mixing and the solar and atmospheric neutrino deficits},
\plb{349}{137}{1995};\\
K. Kang, J. E. Kim and P. Ko, {\it A simple modification of the maximal
mixing scenario for three light neutrinos}, \zp{72}{671}{1996};\\
E. Ma, {\it Plato's fire and neutrino mass matrix},
\mpl{17}{2361}{2002};\\
W. Grimus and L. Lavoura, {\it A discrete symmetry group for
maximal atmospheric neutrino mixing}, \plb{572}{189}{2003};\\
K.-Y. Choi, Y. Kajiyama, J. Kubo, and H. M. Lee, {\it Double
suppression of FCNCs in supersymmetric models},
\prd{70}{055004}{2004}.

\end{thebibliography}
\end{document}